\documentclass[12pt]{article}
\pdfoutput=1
\usepackage{amsmath,amsfonts,sint,epsfig,cite,graphics}
\usepackage{amssymb}
\usepackage{mathrsfs} 
\usepackage{amsbsy,color}
\usepackage{latexsym}
\usepackage{graphicx}
\usepackage{psfrag}
\usepackage{multirow}
\usepackage{booktabs}
\usepackage{url}
\usepackage{dcolumn}
\usepackage[T1]{fontenc}

\newcommand{\be}{\begin{equation}}
\newcommand{\ee}{\end{equation}}
\newcommand{\bea}{\begin{eqnarray}}
\newcommand{\eea}{\end{eqnarray}}
\newcommand{\bes}{\begin{eqnarray}}
\newcommand{\ees}{\end{eqnarray}}
\newcommand{\ba}{\begin{array}}
\newcommand{\ea}{\end{array}}

\newcommand{\eq}[1]{eq.~(\ref{#1})}

\newcommand{\fig}[1]{Fig.~\ref{#1}}
\newcommand{\Fig}[1]{Figure~\ref{#1}}
\newcommand{\sect}[1]{Sect.~\ref{#1}}

\newcommand{\tab}[1]{Table~\ref{#1}}

\newcommand{\Ref}[1]{Ref.~\cite{#1}}
\newcommand{\Refs}[1]{Refs.~\cite{#1}}

\newcommand{\chitop}{\chi_\mathrm{top}}
\newcommand{\tauint}{\tau_\mathrm{int}}
\newcommand{\tauexp}{\tau_\mathrm{exp}}

\newcommand{\SW}{S_\mathrm{W}}

\newcommand{\Pacc}{P_\mathrm{acc}}
\newcommand{\chiglob}{\chi^\mathrm{glob}}
\newcommand{\chicorr}{\chi^\mathrm{corr}}

\newcommand{\nf}{N_\mathrm{f}}
\newcommand{\rmd}{\mathrm{d}}
\newcommand{\rme}{\mathrm{e}}
\newcommand{\rmO}{\mathrm{O}}

\newcommand{\fm}{\mathrm{fm}}

\begin{document}

\begin{titlepage}

\begin{flushright}
\small{
DESY 14-101 \\
SFB/CPP-14-29 \\
}
\end{flushright}

\begin{center}
{\Large\bf
Topological susceptibility and the sampling of field space in $N_\mathrm{f}=2$
lattice QCD simulations
}
\end{center}
\vskip 0.35cm
\vbox{
\centerline{
\epsfxsize=2.8 true cm
\epsfbox{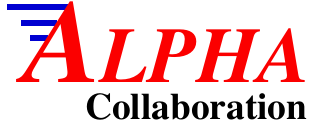}}
}
\vskip 0.1cm
\begin{center}
{
Mattia Bruno,
Stefan Schaefer and
Rainer Sommer
}
\vskip 0.5cm
{
  John von Neumann Institute for Computing (NIC), DESY\\
  Platanenallee~6, D-15738 Zeuthen, Germany
\vskip 2.0ex
}
\vskip 0.5cm
{\bf Abstract}
\vskip 0.1ex
\end{center}

We present a measurement of the topological susceptibility in two flavor QCD.
In this observable, large autocorrelations are present and also sizable
cutoff effects have to be faced in the continuum extrapolation. Within the 
statistical accuracy of the computation, the result agrees with the expectation 
from leading order chiral perturbation theory.


\vfill
\eject

\end{titlepage}

\section{Introduction}

The topology of the gauge fields plays an important role in the understanding
of the low energy structure of QCD. A prominent example is the Witten--Veneziano
relation where the topological susceptibility of pure Yang--Mills theory is 
linked to the mass of the  $\eta'$ meson\cite{Witten:1979vv,Veneziano:1979ec}.
The object of this paper is the topological susceptibility  $\chitop$ in two flavor QCD,
a quantity which vanishes in the chiral limit but whose value is non-zero for
finite sea quark masses.

Precise measurements of $\chitop(m_\pi)$ are a challenge for the lattice. In 
particular significant effort went into finding a suitable definition of the
topological charge on a discrete space time. Also large 
autocorrelations in the numerical simulations and sizable 
cutoff effects make the determination difficult.

These issues have a long history which we can only sketch briefly. The naive
``field theoretical'' definition of the topological charge density
\begin{equation}
q(x)= - \frac{1}{32\pi^2}\, \epsilon_{\mu\nu\rho\sigma}\, \mathrm{tr} \{ F_{\mu\nu}(x) \,  F_{\rho\sigma}(x) \} \,,
\end{equation}
using some discretization of the  field strength tensor $F$ is known to lead
to a topological susceptibility $\chitop$ 
\begin{equation}
\chitop=\frac{1}{V} \int dx\,\int d y\, \langle\,q(x)\, q(y)\,\rangle \,,
\label{eq:chitop}
\end{equation}
which suffers from non-integrable short-distance singularities. 
This has led to a vast array of operational procedures based on cooling and
link smearing which remove the ultra-violet fluctuations of the gauge fields,
but whose behavior towards the continuum limit is not well understood.
However, a number of practical definitions with a well-defined continuum limit
are now available, notably based on the properties of  chiral Dirac operators and the
index theorem\cite{Hasenfratz:1998ri} or ratios of certain fermionic correlation
functions\cite{Giusti:2001xh,Giusti:2004qd,Luscher:2004fu}. 

In this study, we adopt the definition of the topological charge through the
Wilson flow\cite{Luscher:2010iy}, which provides a smoothing of the gauge fields
with a smearing radius that is kept constant in physical units as the continuum 
is approached. Numerical support for a common continuum limit of this 
definition of the susceptibility and the method of 
Refs.~\cite{Giusti:2001xh,Giusti:2004qd,Luscher:2004fu} in pure gauge theory has 
been given in Ref.~\cite{Luscher:2010ik}.

While the computation of the topological charge using the Wilson flow requires
only a moderate numerical effort, the simulations are still expensive due
to large autocorrelation times in the topological charge. In particular if 
periodic boundary conditions are adopted, the global topological charge suffers
from a severe critical slowing down\cite{Schaefer:2010hu} and a determination
of the susceptibility becomes practically impossible. The particularly poor
scaling of the autocorrelation times due to the topological charge can be 
solved by using open boundary conditions in time\cite{Luscher:2010we,Luscher:2011kk},
however, the smooth fields in the construction of the topological charge
are still slow in the Monte Carlo evolution.

Studying the effect of dynamical fermions on the susceptibility using lattice QCD
has a long tradition, but even a decade ago, the situation was less than clear.
While the Sesam collaboration did not find a convincing suppression of the susceptibility
\cite{Bali:2001gk}, the UKQCD collaboration did observe some effect\cite{Hart:2001pj}.
Later, the importance of cut-off effects in this observable was discussed for
simulations with Wilson fermions\cite{Hart:2004ij} and staggered quarks\cite{Bernard:2003gq,Bazavov:2010xr}.
After taking the continuum limit, a behavior compatible with the 
continuum expectation from Chiral Perturbation Theory\cite{Leutwyler:1992yt}
was found. More recently, the suppression of the susceptibility with the
quark mass in two flavor QCD was also demonstrated  in \Refs{Cichy:2013rra}
and \cite{Chiu:2011dz} using Wilson fermions with a twisted mass and domain wall
fermions, respectively.\footnote{In \Ref{Cichy:2013rra}, 
there is a single lattice spacing for the 2-flavor results. For $N_\mathrm{f}=$2+1+1
three lattice spacings are present but the continuum extrapolation assumes 
the susceptibility to vanish in the chiral limit at fixed lattice spacing, see our 
discussion in \sect{s:cont}.}

Due to the limitations by the statistical error and the discretization effects,
the scope of this study is twofold: On the one hand, we want to present the
measurement of  $\chitop(m_\pi)$ as well as it can be made with currently
available ensembles and present evidence for the proper suppression of the 
topological charge fluctuations with the quark mass.  On the other hand, studying the autocorrelations
in a set of gauge field configurations which are used in many computations
serves as a crucial input in the error analysis of these
computations\cite{Schaefer:2010hu}.  
We will also have a first look at the impact of
open boundary conditions in time in simulations including dynamical fermions.

The paper is therefore organized as follows: In \sect{sec:2} the observables
are defined and an overview of the ensembles used in the current study is given.
We examine the autocorrelations in observables constructed from smoothed gauge fields
and give new estimates of the largest exponential autocorrelation times.
In \sect{sec:3}, after the separation of the gauge fields into topological sectors is presented
for the two flavor theory, we then turn to the computation of the susceptibility,
the discussion of its discretization effects and a comparison of the continuum
limit of $\chitop(m_\pi)$ with Chiral Perturbation Theory.

\section{Setup of the calculation\label{sec:2}}

The Wilson flow in the space of lattice gauge fields $U(x,\mu)$ is defined by the equations
\begin{equation}
\partial_t V_t(x,\mu) = -g_0^2 \{ \partial_{x,\mu} \SW (V_t) \} V_t (x,\mu) 
\,, \quad V_t(x,\mu)\big|_{t=0} = U(x,\mu)\,,
\end{equation}
where the action $\SW$ is the standard Wilson action and $V_t$ are the smoothed link
variables at flow-time $t$. It was introduced
in \Ref{Luscher:2010iy} and we follow the conventions  adopted there.
In particular we will look at the topological charge density $q(x,t)$ 
\begin{equation}
q(x,t)=-\frac{1}{32\pi^2}\, \epsilon_{\mu\nu\rho\sigma}\, \mathrm{tr} \{ G_{\mu\nu}(x,t) \,  G_{\rho\sigma}(x,t) \} \,,
\label{def:tcharge_density}
\end{equation}
with a clover-type discretization of the  field strength tensor 
$G_{\mu\nu}(x)$ constructed from links $V_t$.
Furthermore the energy density constructed from these links will be considered
\begin{equation}
E(t)=-\frac{1}{2V}\sum_x  \mathrm{tr} \{ G_{\mu\nu}(x,t) \,  G_{\mu\nu}(x,t) \} \,.
\label{def:energy}
\end{equation}

The quantities defined through the flow need to be evaluated at a fixed, physical
value of $t$. Since it provides a smoothing of the fields over a radius $r=\sqrt{8t}$,
it is natural to use the reference scale $t_0$ introduced in the original 
paper\cite{Luscher:2010iy} through
\begin{equation}
t^2 \langle E(t) \rangle \big|_{t=t_0}  = 0.3 \,.
\label{eq:t0}
\end{equation}
Recently it has been argued that cutoff effects in some quantities can be reduced by 
considering the flow at a larger time $t_1$, where  the right hand side of \eq{eq:t0}
is replaced by $2/3$ \cite{Sommer:2014mea}. We will give results in units of $t_0$ and $t_1$. To convert to physical units the reader may use 
$t_0 = 0.0236\,\fm^2$ \cite{Bruno:2013gha} and
$t_1 = 0.061\,\fm^2$, both obtained for $\nf=2$ from fixing the kaon decay constant
and the pion mass, $m_\pi$, to experiment. 
Their errors are negligible for our present purposes.
While these numbers refer to the physical point, in 
the rest of this paper we always use the scales $t_0(m_\pi),\, t_1(m_\pi)$ as obtained at the simulated
values of $m_\pi$.

The topological susceptibility derived from the charge density \eq{def:tcharge_density}
does not require renormalization. While its continuum limit does not depend on the flow time $t$,
discretization effects will  be influenced by its choice. However, we will
see below that choosing either $t=t_0$ or $t=t_1$ does not have a sizable effect on $a^4\chitop$.

Starting from the topological charge density defined in \eq{def:tcharge_density}, 
the topological charge can be directly obtained
\begin{equation}
Q(t)=a^4 \sum_x q(x,t)\,.
\end{equation}
In periodic boundary conditions, the susceptibility is then simply $\chitop=\langle Q^2 \rangle/V$.

\subsection{Ensembles}

The set of ensembles used for our study was generated within the CLS effort.\footnote{\url{https://twiki.cern.ch/twiki/bin/view/CLS/WebHome}}
Two degenerate light quarks have been simulated using $\mathrm{O}(a)$-improved Wilson
fermions and Wilson gauge action. The ensembles are listed in \tab{tab:ensembles}: three
lattice spacings and a range of  pion masses  are available with all lattices fulfilling $m_\pi\,L>4$.
\begin{table}[th]
\centering
\begin{tabular}{llcccccccc}
\toprule
$\beta$ & ID & $T/a$ & $L/a$ & $\kappa$ &  $m_\pi$ [MeV] & $R$ & $\Pacc$ & MDU & stat$/\tau_{exp}$ \\
\midrule
5.2 & A2  & 64 & 32 & 0.13565 & 630 & 0.4 & 0.96 & 2953 & \phantom{1}76 \\ 
    & A3  & 64 & 32 & 0.13580 & 490 & 0.4 & 0.92 & 2965 & \phantom{1}76 \\ 
    & A4  & 64 & 32 & 0.13590 & 380 & 0.4 & 0.86 & 2989 & \phantom{1}76 \\ 
    & A5  & 64 & 32 & 0.13594 & 330 & 1.0 & 0.93 & 4004 & 102 \\ 
    & B6  & 96 & 48 & 0.13597 & 280 & 1.0 & 0.99 & 1272 & \phantom{1}33 \\ 
    & oB6 & 192& 48 & 0.13597 & 280 & 1.0 & 0.85 & 1000 & \phantom{1}26 \\
\midrule                                     
5.3 & E5g & 64 & 32 & 0.13625 & 430 & 0.4 & 0.84 & 5906 & 106 \\ 
    & F6  & 96 & 48 & 0.13635 & 310 & 0.4 & 0.88 & 1772 & \phantom{1}32 \\ 
    & F7  & 96 & 48 & 0.13638 & 260 & 0.4 & 0.86 & 3550 & \phantom{1}63 \\ 
    & G8  & 128& 64 & 0.13642 & 190 & 1.0 & 0.99 & 1756 & \phantom{1}31 \\ 
\midrule                                     
5.5 & N6  & 96 & 48 & 0.13667 & 340 & 1.0 & 0.83 & 8040 & \phantom{1}74 \\ 
    & O7  & 128& 64 & 0.13671 & 260 & 1.0 & 0.84 & 3920 & \phantom{1}36 \\

\bottomrule
\end{tabular}
\caption{Overview of the ensembles used. We give the parameters
of the action $\beta$ and $\kappa$, the temporal and spatial extent
of the lattice and the pion mass. For the simulations done with the
DD-HMC algorithm, the ratio of active links, $R$, with which the $\tauint$ are
scaled throughout the paper, is below one. The last two columns
give the total statistics in molecular dynamics units (MDU) 
and how this compares to the $\tauexp$ estimated
in \eq{eq:tauexp}.
\label{tab:ensembles}}
\end{table}

The majority of the ensembles was generated with the DD-HMC algorithm\cite{Luscher:2005rx}, 
while the  HMC algorithm with Hasenbusch preconditioning\cite{Hasenbusch:2001ne}
has been employed for those with smaller pion masses\cite{Marinkovic:2010eg}. The ensemble
where open boundary conditions in time have been imposed has been generated with 
the openQCD code\cite{Luscher:2012av}.
The details of most of the simulations  can be found in \Ref{Fritzsch:2012wq}.

Since a significant part of this text deals with autocorrelations, it should be noted that 
algorithms and their parameters influence them. The different ways the fermion determinants are
treated in the three algorithmic setups will certainly matter once accuracies are high. At
the current level of precision, we only correct for the inactive links during the trajectories
of the DD-HMC algorithm by scaling the molecular dynamics time with $R$, the fraction of active links.
In the high statistics  pure gauge theory  study of \Ref{Schaefer:2010hu} this has proven 
to correctly compensate for their effect.
We will also scale autocorrelation times with the acceptance rate $P_\mathrm{acc}$ in the final
scaling formula.

\subsection{Autocorrelations}

Large autocorrelations are a well-known problem in lattice simulations. They are particularly
pronounced in quantities constructed from the fields smoothed by the Wilson flow.
In these quantities, a critical slowing down governed by a dynamical critical exponent of 
$z=2$ is expected\cite{Luscher:2011qa}. To compare simulations at different lattice spacings, 
we will therefore scale the Monte Carlo time with the dimension two quantity $t_0$ measured
on the respective ensemble.

An additional problem is posed by the freezing of the topological charge as the continuum 
limit is approached. This occurs with a larger critical exponent -- if it is not exponential --
and makes simulation below $a \approx 0.05$~fm exceedingly difficult.
The problem is not restricted to the topological charge
alone. In general, the associated slow eigenmodes of the Markov chain transition matrix
contribute to the autocorrelation functions of all observables. 
For a correct estimate of the error a careful study of the slowest modes of the
Markov chain is mandatory. 

The basis of the error analysis of Markov Chain Monte Carlo data
is the measurement of the normalized autocorrelation function $\rho_A$
\begin{align}
\rho_A(\tau)&=\frac{\Gamma(\tau)}{\Gamma(0)} \ ;  && 
\Gamma(\tau) =  \langle (A(\tau)-\bar A)  (A(0)-\bar A) \rangle \ ,
\end{align}
with $A(\tau)$ the Monte Carlo time history of the observable $A$ and $\tau$ the 
Monte Carlo time.

From the analysis of the spectral decomposition a recipe to include the coupling 
to the slow modes of the transition matrix in the error analysis has been
given in \Ref{Schaefer:2010hu}: the normalized autocorrelation function  $\rho$
is summed as usual up to a window $W$ and then the contribution of a single 
exponential with time constant $\tauexp$ is attached
\begin{equation}
\tauint = \frac{1}{2} + \sum_{\tau=1}^{W-1} \rho(\tau) + \tauexp \rho(W) \ .
\end{equation}
For this method an estimate of the largest time constant $\tauexp$ in the 
simulation is needed.
Since the dynamics of  parity even and parity odd observables 
decouples\cite{Schaefer:2010hu},
we can restrict ourselves to even ones. Then $Q^2(t)$ is an obvious
candidate to look for the largest time constant, but we also investigate $E(t)$.

\begin{figure}
\centering
\includegraphics[width=0.7\textwidth]{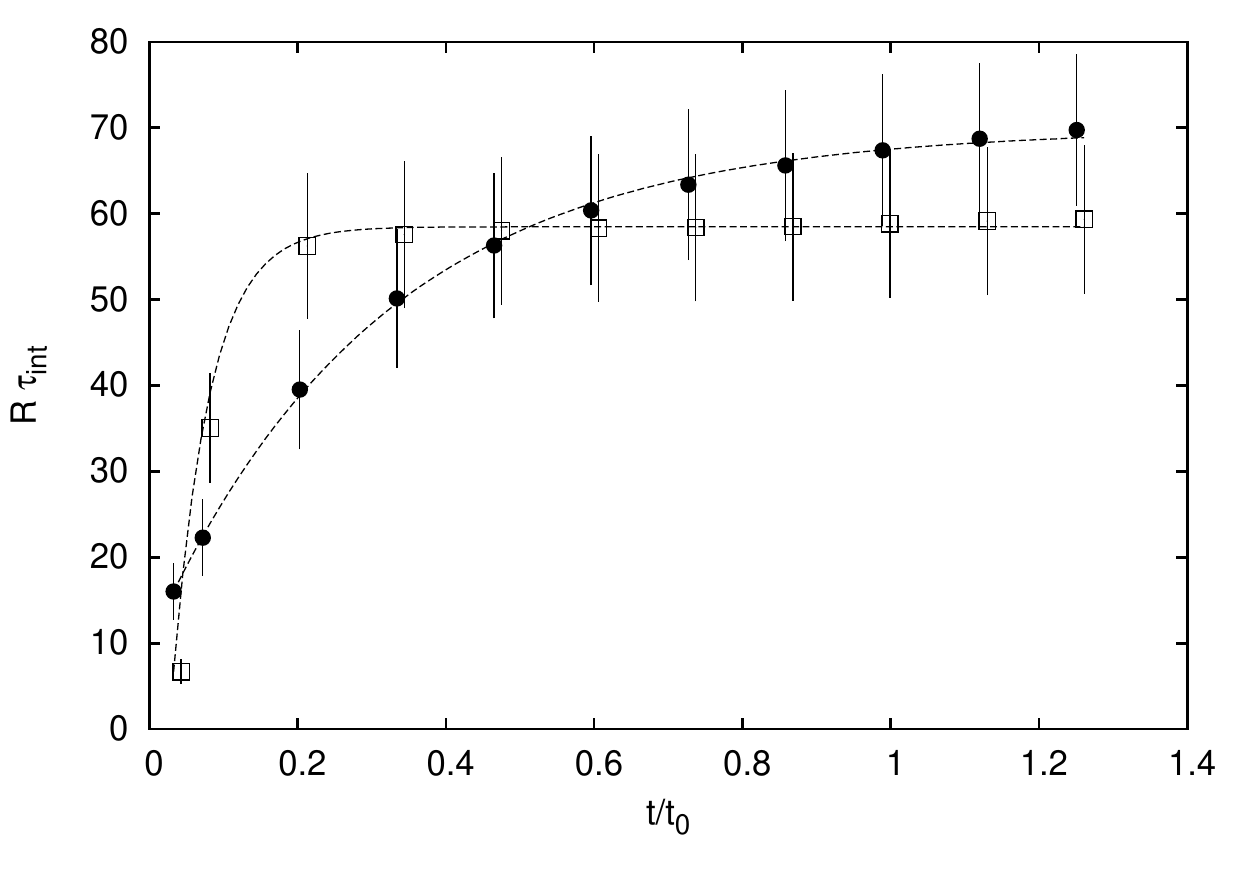}
\caption{Integrated autocorrelation times of $Q(t)^2$ (open) and $E(t)$ (filled) 
from the ensemble N6, see \tab{tab:ensembles}. Points have been shifted for better legibility. 
$\rho(\tau)$ is summed up to a window $W=100$\,MDU, neglecting a potential tail. 
The lines are fits of \eq{e:taut} to the data with with $c_0 \approx 58$\,MDU for $Q(t)^2$ 
and $c_0 \approx 69$\,MDU for $E(t)$.} 
\label{fig:tauint_vs_t}
\end{figure}

For the autocorrelation analysis we have to pick a value of the flow time $t$. In 
 \fig{fig:tauint_vs_t} we present the dependence of $\tauint$ on this parameter and
 find a behavior similar to the one observed in pure gauge theory\cite{Luscher:2012av}:
 the data can be described by 
 \begin{equation}
\tauint(t)=c_0+c_1 e^{-c_2 t}\label{e:taut} \ ,
 \end{equation}
 with the value at $t=t_0$ essentially capturing the asymptotic value. We therefore
 use this value of the flow time and drop the flow time argument in the 
 following, using $Q=Q(t_0)$ and $E=E(t_0)$.

\subsection{Results for autocorrelation times}

\begin{figure}
\centering
\includegraphics[width=0.7\textwidth]{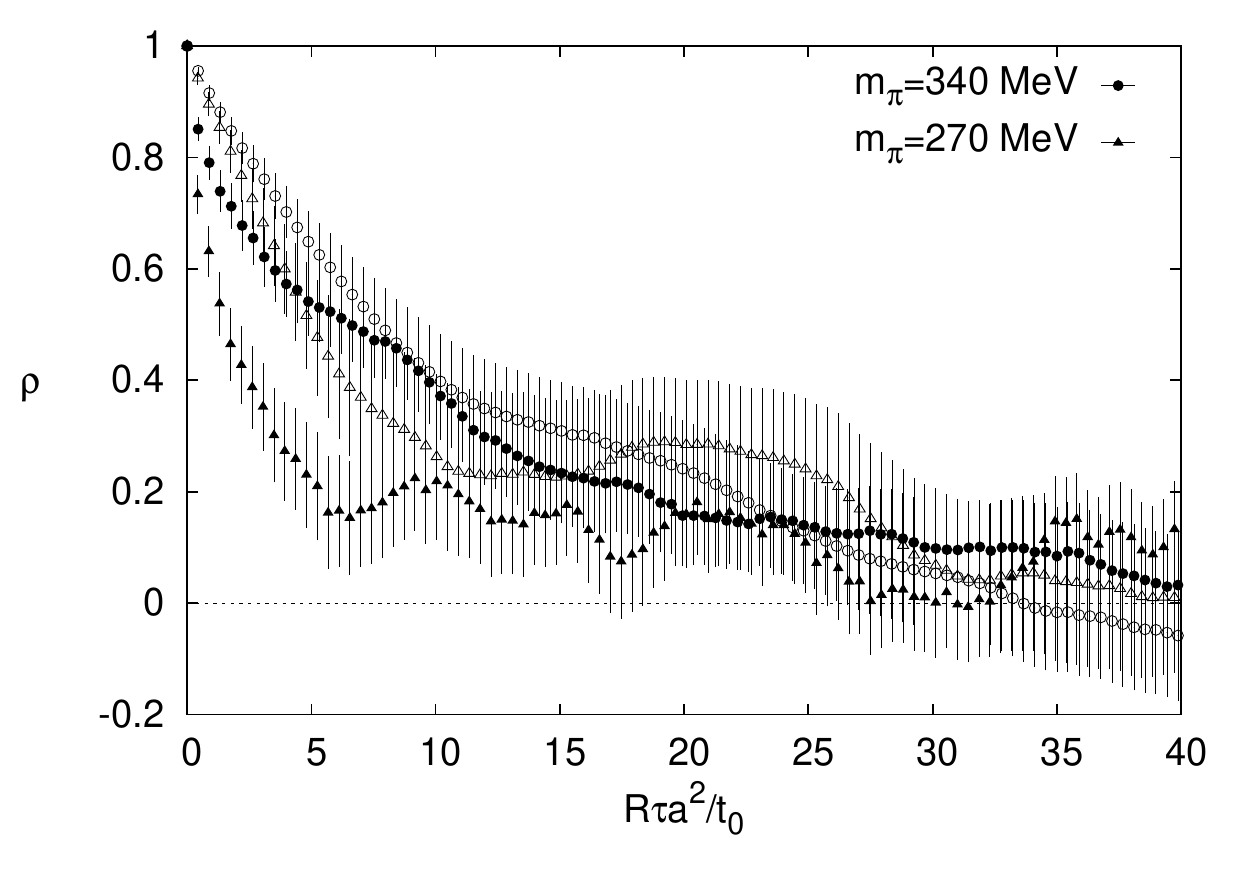}
\caption{Autocorrelation function $\rho(\tau)$ at different pion masses for $\beta=5.5$. For $Q^2$ (filled symbols) there is a 
clear suppression at smaller quark masses, whereas $E$ (open symbols) does not show an effect beyond
the statistical accuracy.}
\label{fig:autocorr-2}
\end{figure}

Autocorrelation functions depend on the observable, the lattice spacing and the quark mass.
To quantify the dependence on the physics parameters, we give examples of autocorrelation functions for
two values of the quark mass in \fig{fig:autocorr-2}. 
The data at $\beta=5.5$ shows a suppression with the quark mass of 
 $\rho_{Q^2}$, as also found in \Ref{Chowdhury:2012qm}, whereas $\rho_{E}$ is not affected beyond statistics.
 This can be understood from the fact that 
 the topological susceptibility goes to zero in the chiral limit.
The narrower distributions can be sampled in fewer steps leading to faster decorrelation in Monte Carlo time.

\begin{figure}
\centering
\includegraphics[width=0.7\textwidth]{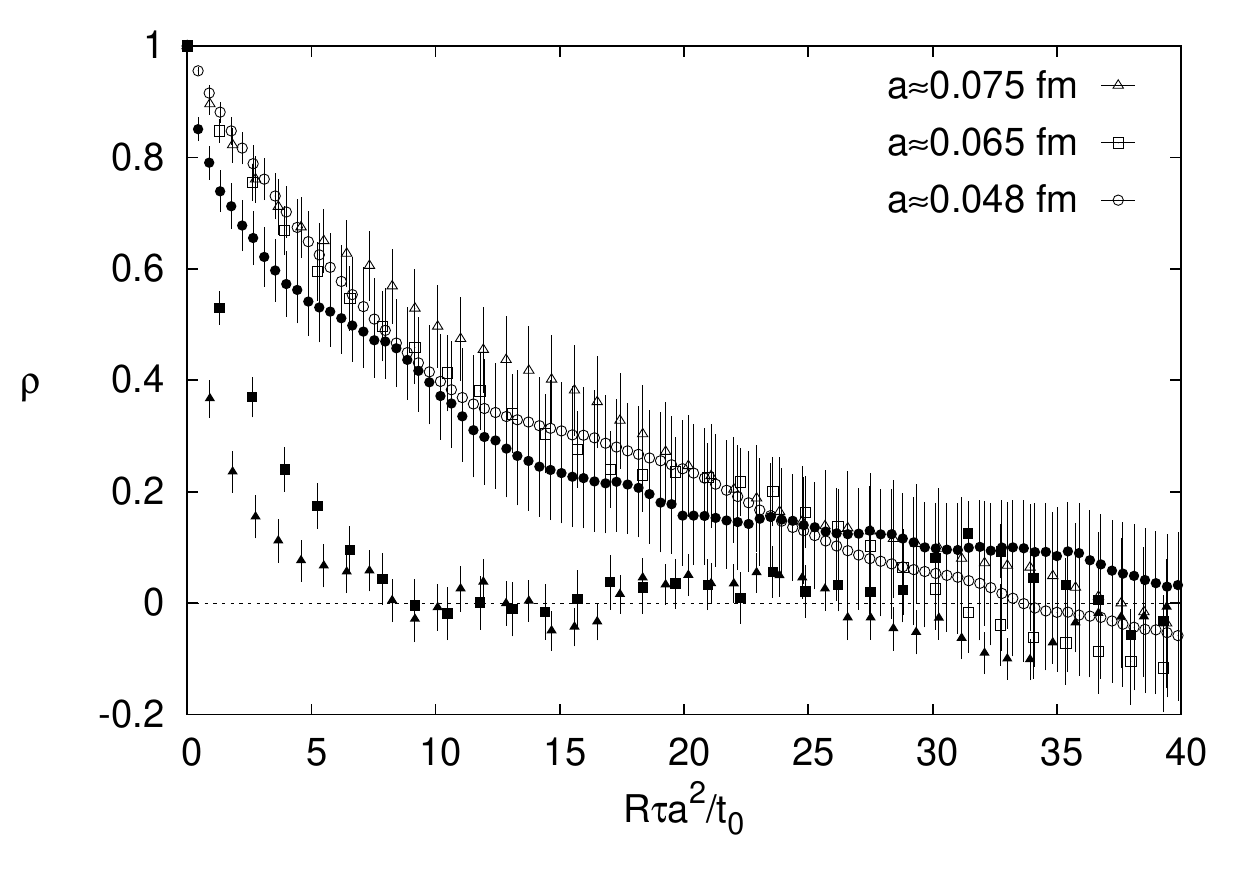}
\caption{Autocorrelation function for different values of the lattice spacing.
Open symbols correspond to $E$ and filled to $Q^2$.
$\rho_E$ shows the expected scaling, whereas $Q^2$ is significantly 
faster on the coarser lattices.
\label{fig:autocorr-1}
}
\end{figure}

 \Fig{fig:autocorr-1} shows the dependence of the
autocorrelation functions on the lattice spacing. While the rescaling of the
Monte Carlo time with $t_0$ confirms the scaling properties of $\rho_E$ already 
observed in pure gauge theory, the topological charge decorrelates much faster
on the coarser lattices. On our finest lattice spacing, the autocorrelation functions
of $E$ and $Q^2$ are on top of each other and we expect the topological charge
to dominate on finer lattices. This qualitative behavior is again similar to the one
in pure gauge theory with periodic boundary conditions\cite{Luscher:2010we},
only that it happens at smaller lattice spacing --- not a surprise
in  light of the suppression of autocorrelations in $Q^2$ with the quark mass.

\begin{table}[tbh]
\centering
\begin{tabular}{@{\extracolsep{0.5cm}}ccc}
\toprule
ID & $R \tauint(E)$ & $R \tauint(Q^2)$ \\
\midrule

A2  & 55(22) & 6(1)  \\ 
A4  & 42(16) & 5(1)  \\ 
A5  & 49(18) & 4(1)  \\ 
B6  & 45(21) & 4(1)  \\ 
oB6 & 30(14) & ---   \\
\midrule

E5g & 47(15) & 12(3)  \\ 
F6  & 36(15) & 10(3)  \\ 
F7  & 83(35) & 11(3)  \\ 
G8  & 57(26) & 4(1)   \\ 

\midrule

N6 & 105(38) & 100(41) \\ 
O7 & 102(44) & 52(21) \\

\bottomrule
\end{tabular}
\caption{Integrated autocorrelation 
times measured on our set of ensembles where sufficient 
statistics is available. }
\label{tab:tauint}
\end{table}

The results of the integrated
autocorrelation times (summed up to $R\tau \approx 40 t_0 / a^2$) for $Q^2$ and $E$ are given in \tab{tab:tauint}.
We also tried single exponential fits to the  various $\rho_E$ in order to 
find $\tauexp$, the largest time constant of the Monte Carlo chain.
It turns out that the $\rho(\tau)$ are well described  by this ansatz over 
essentially the full range of $\tau$. 

Since $E$ has the largest autocorrelations in our range of lattice spacings
and its autocorrelation function is approximately an exponential function,
we can identify  $\tauint(E)$ with $\tauexp$ at the current level of accuracy.
This leads to a parametrization valid for $\beta\in [5.2,5.5]$
\begin{equation}
R \tauexp(\beta) = 11.3(1.8) \dfrac{t_0}{a^2} \,.
\label{eq:tauexp}
\end{equation}

In an earlier publication\cite{Schaefer:2010hu}, an estimate of $\tauexp(\beta)$ has been given
based on the scaling of $\tauint(Q^2)$ in pure gauge theory normalized to the
value of the E5 lattice. Since the topological charge is not the slowest 
quantity in this region, this is overly pessimistic for $\beta=5.5$, where it
gives twice the $\tauint(Q^2)$ observed and overly optimistic at $\beta=5.2$.

Since the autocorrelations of the topological charge show a scaling compatible with 
what has been found in pure gauge theory, i.e. $\tauint(Q^2)\propto a^{-5}$,
we expect that for our range of quark masses and  $\beta>5.5$ the charge will 
dominate  the scaling of $\tauexp$.

\subsubsection{Open boundary conditions}
The use of open boundary conditions is expected to accelerate the sampling of the
topological charge, which at fine lattice spacings is expected to be the slowest
observable. However, we observe that this is not the case for our lattices,
in particular at $\beta=5.2$ where oB6 has been simulated. Here the bulk tunneling
dominates over the effect of the open boundaries.

Within the accuracy of the result for $\tauint(E)$ reported in \tab{tab:tauint},
no advantage in terms of autocorrelations is visible between B6 and oB6.
For this comparison, we compute the energy density on the time slices in 
$x_0/a \in [20,171]$ in which the effect on the observable from the boundaries is negligible.

With open BC the notion of a global topological charge is lost, as will be
discussed in the next section, therefore we do not compute $\tauint(Q^2)$ for oB6.
For the alternative way to extract the susceptibility discussed below,
we find comparable autocorrelation times for the two ensembles.

\section{Results\label{sec:3}}

From the discussion in the previous section it should be clear that while we
are confident that the statistics is sufficient for having control over the
statistical errors, the  accuracy of the susceptibility which can be reached from these
ensembles only allows for qualitative statements. It is certainly not on par
with the precision reached in pure gauge theory.

We will start by discussing the separation of topological sectors
as the continuum is approached. In some sense, this is a measure 
for the physics associated with topology being realized already 
at a finite lattice spacing. It also is the reason
for the Hybrid Monte Carlo algorithm having difficulty to 
decorrelate the topological charge.

Once this is established, we turn to the topological susceptibility, studying
systematic effects between different ways of constructing it. This will
be important for its determination imposing open boundary conditions
in time, where the global topological charge is no longer a suitable
observable.

\subsection{Separation between the sectors}

With periodic boundary conditions field space is disconnected in the
continuum gauge theory. This property is obfuscated by ultra-violet fluctuations
of the gauge fields and for a long time whether and in which way these sectors emerge as the lattice spacing decreases has been unclear.
Using the smoothing provided by the Wilson flow, 
the emergence of the topological sectors in the continuum limit
can finally be understood \cite{Luscher:2010we}. 
For convenience we here repeat the arguments in \cite{Luscher:2010we} 
in a rather explicit manner. 

First one notices that it was shown long ago, that the space of 
lattice gauge fields $U$ separates into disconnected sectors provided the
gauge fields $U$ are smooth enough, which means that 
\be
    s_p(U) =  \mathrm{Re}\,\mathrm{tr} \{ 1-U(p) \}
\ee
satisfies 
\be
    s_p(U) < s_\mathrm{cut} = 0.067 \,,
    \label{e:bound}
\ee
for all plaquettes $p$ on the lattice
\cite{Phillips:1986qd,Luscher:1981zq}.\footnote{
For attempts to construct gauge actions which satisfy the bound automatically 
{\em and} can be used in numerical simulations, see \cite{Bietenholz:2005rd}
while for related questions of universality we refer 
to \cite{Bietenholz:2010xg}.}

Secondly, the Wilson flow defines a map
$U\equiv V_0 \leftrightarrow V_{t}$ of the gauge fields in our 
path integral to smoothed fields $V_t\,,\; t>0$. Since this map is 
one-to-one, a classification of $V_t$ in sectors is also one
of the original fields $U$. Fields $V_t$
are smooth since the total (global) action 
$S_\mathrm{W}(V) = \sum_p \frac{1}{g_0^2} s_p(V)$ is decreased by the flow,
$S_\mathrm{W}(V_t)<S_\mathrm{W}(V_{t'})$ for $t>t'$. Of course, this does
not mean that the local bound \eq{e:bound} will be satisfied for $V_t$
at a given $t$. However, if the number of plaquettes which violate
$s_p(V_t) < s_\mathrm{cut}$ decreases strongly as one approaches $a\to 0$, 
we can say that sectors develop dynamically in the continuum limit.

Thirdly, in \Ref{Luscher:2010we} violations of the bound \eq{e:bound} are
investigated numerically using the probability that one plaquette violates the bound
\be
   P_t(s_\mathrm{cut}) = 
   \frac{1}{6V} \langle \sum_p \theta(s_\mathrm{cut} - s_p(V_t)) \rangle \,, 
\ee
It finds that in pure gauge theory with the Wilson action 
$P_t(s_\mathrm{cut}) \sim a^{10}$, which also suggests that in a fixed volume
the probability for any plaquette violating the bound decreases as $a^6$.

\begin{figure}
\begin{center}
\includegraphics[width=0.7\textwidth]{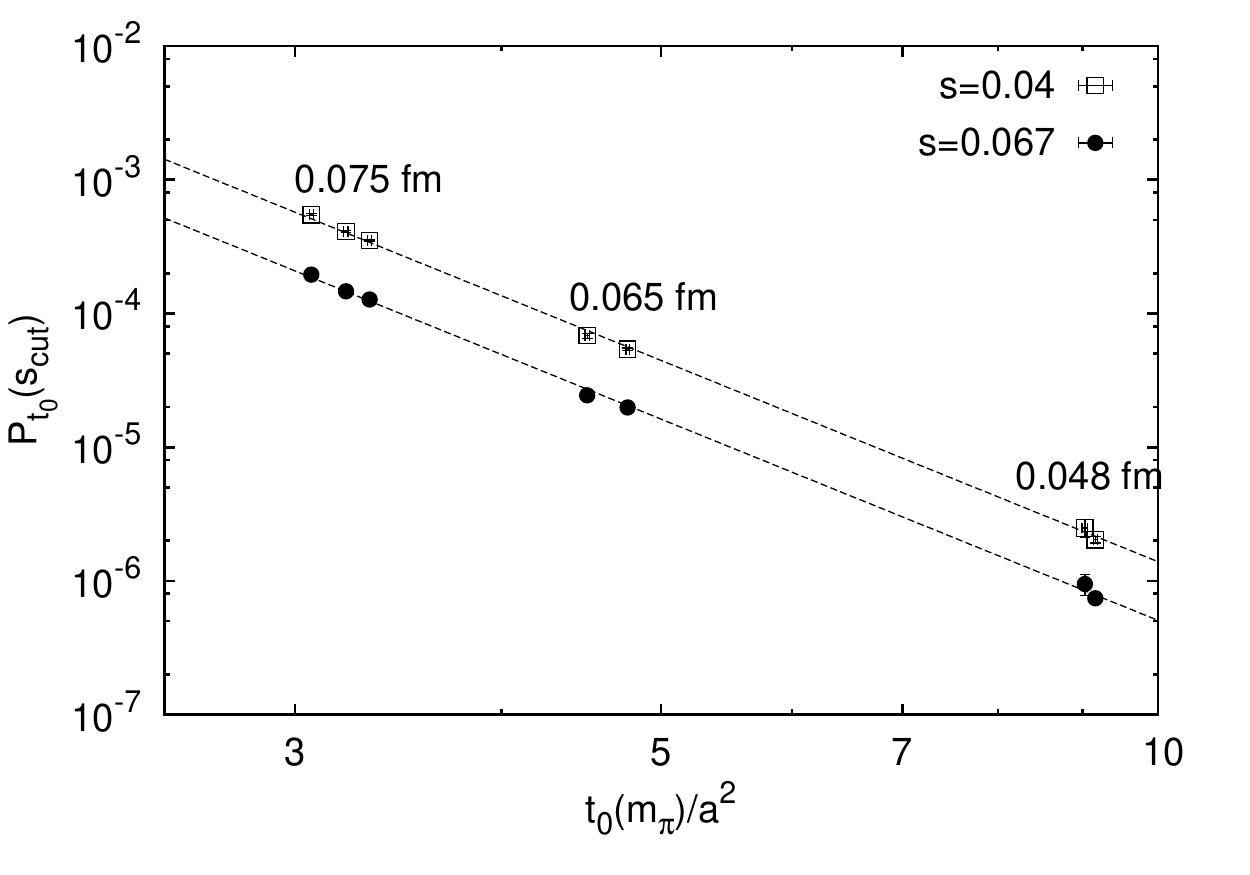}
\end{center}

\caption{The probability $P_{t_0}(s_\mathrm{cut})$ of finding a plaquette value 
$s_p(V_t)\geq s_\mathrm{cut}$ as a
function of the lattice spacing. The dotted lines represent a power law
decay with $P_{t_0}(s_\mathrm{cut})\propto a^{10}$. \label{fig:sector}}
\end{figure}

As can be seen in \Fig{fig:sector}, we find the same power law behavior in
 our two flavor simulations. While two values of $s_\mathrm{cut}$ are shown, 
 the scaling does not depend significantly on $s_\mathrm{cut}$ in the investigated range
 up to $s_\mathrm{cut}=0.1$ and the exponent
is compatible with the pure gauge theory result of \Ref{Luscher:2010we}.
It is remarkable that the effect of the quark mass on $P_{t_0}(s_\mathrm{cut})$ is completely compensated
by its effect on $t_0$ and we observe universal scaling curves for each value of $s_\mathrm{cut}$.
At a fixed $\beta$, smaller quark masses lead to a stronger separation of the sectors. In particular the presence of fermions does not alter the picture
of \Ref{Luscher:2010we}.

\subsection{Computation of the topological susceptibility}
With periodic boundary conditions for the gauge fields,
\begin{equation}
\chiglob=\frac{1}{V} \langle Q^2 \rangle
\label{eq:chiglob}
\end{equation}
is the natural estimator for the topological susceptibility.
With open boundary conditions in time, the same quantity would show large
finite volume effects.\footnote{With significant statistical uncertainties, \Ref{Chowdhury:2013mea} does not resolve
finite $T$ effects in a comparision of $\chitop$ from open and periodic boundary conditions at flow
time $t=t_0$. At smaller $t$, however, deviations are seen.} 
Therefore we rewrite \eq{eq:chitop}
restricting the time separation between the two densities to an upper bound $r$
\begin{equation}
\chicorr(r)= \frac{a^2}{TL^3}\sum_{z_0=0}^{T-1} \sum_{x_0=-r}^{r} \, \langle \bar q(z_0) \bar q(x_0+z_0) \rangle \ \ ;
\ \  \bar q(x_0) =  a^3 \sum_{\bf x}\, q(x_0,{\bf x}) \,,
\label{eq:chicorr}
\end{equation}
with 
$\chicorr(T/2\, - a) + \,  \frac{a^2}{TL^3}\langle \bar q(0) \bar q(T/2) \rangle $
giving the full formula for periodic boundaries with even $T/a$. 
The correlator $\langle \bar q(z_0) \bar q(x_0+z_0) \rangle$ will be studied below.
As we will see, it falls off quickly for large $x_0$ such
that the sum can be truncated at moderate $r$. 

To avoid finite $T$ effects, all temporal arguments need to be sufficiently far
away from the boundaries, i.e., $x_0$, $T-x_0$, $x_0+z_0$, $T-(x_0+z_0)$ need to be
large in units of the inverse mass of the lightest state with vacuum quantum numbers.

\begin{figure}
\centering
\includegraphics[width=\textwidth]{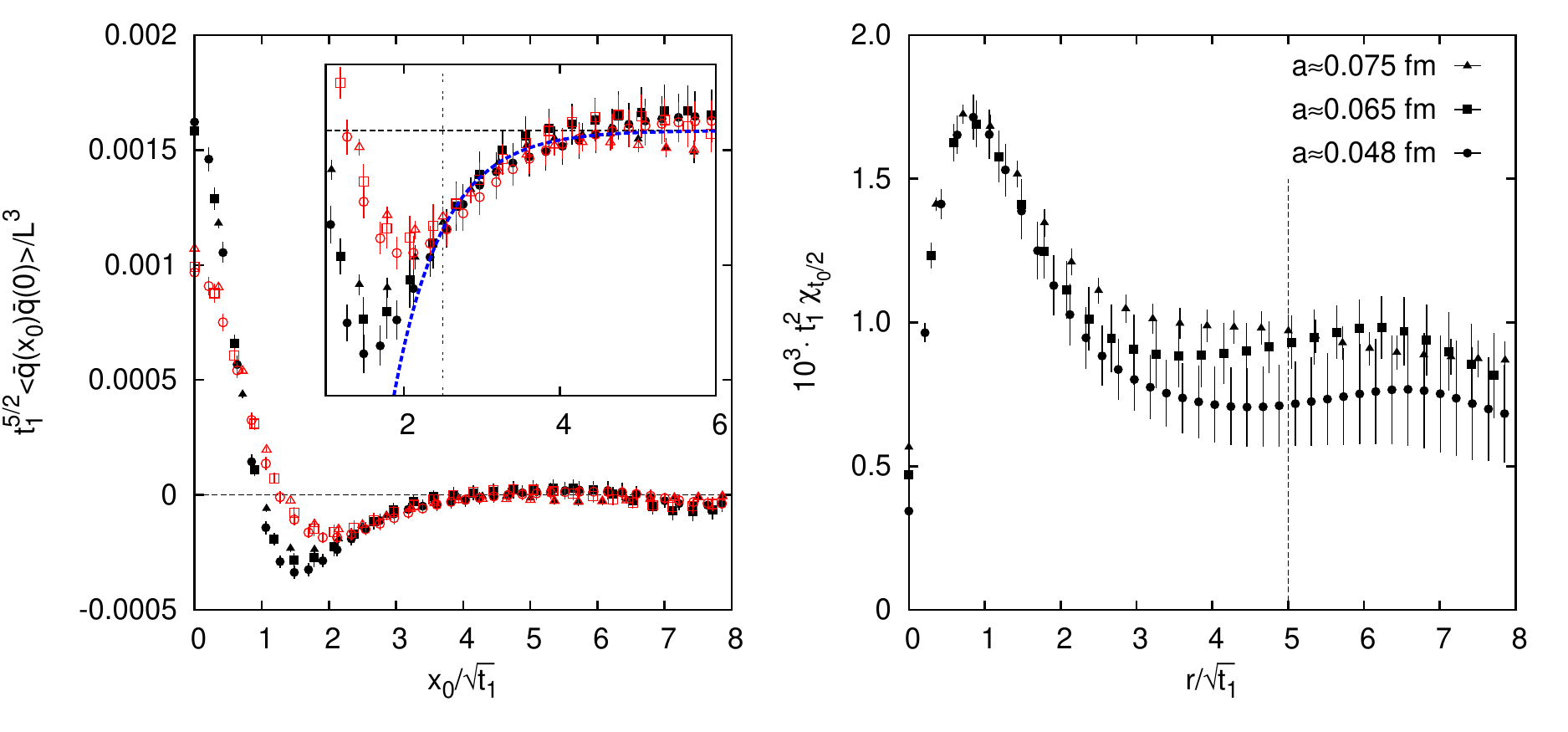}
\caption{Determination of the susceptibility from the
correlation function of the time slice summed charge at
$t=t_0$ (open symbols) and $t=t_0/2$ (filled symbols); the scales $t_0$ and $t_1$  are taken
at the quark mass of the ensemble.
In both plots the ensembles are A5, F6 and N6. For the 
time-slice correlation functions in the left panel, the long
distance behavior can be described by a single exponential
decay as discussed in the text. In the right
panel, the sum up to a window $r$ is represented, with the vertical line the
one used in our analysis.
\label{fig:chit0}}
\end{figure}

In order to study the effect of the truncation in \eq{eq:chicorr}, we plot the  correlator between 
the slice charges for several values of the lattice spacings on the periodic lattices
in the left panel of \fig{fig:chit0}. This correlation function
is positive for small distances, then turns negative and
approaches zero for $x_0\to \infty$.  The sum over $x_0$ as a function of the upper bound is 
depicted on the right panel. As we can see,  
for our statistics a plateau is reached around $r/\sqrt{t_1}=5$. 
This holds for all our ensembles. We therefore cut the summation
at this point; the results are collected in \tab{tab:topsusc}.

\begin{table}[tbh]
\centering
\begin{tabular}{@{\extracolsep{0.5cm}}cccccc}
\toprule
ID & $t_1^2 \chiglob_{t_0/2}$ & $t_1^2 \chicorr_{t_0/2}$ & $t_0^2 \chicorr_{t_0/2}$ & $t_1^2 \chiglob_{t_0}$ & $t_1^2 \chiglob_{t_1}$\\
\midrule

A2  & 1.48(14) & 1.46(11) & 0.27(02) & 1.67(17) & 1.76(17) \\ 
A4  & 1.06(10) & 1.10(08) & 0.19(01) & 1.21(12) & 1.26(12) \\ 
A5  & 0.88(07) & 0.97(06) & 0.17(01) & 0.94(07) & 1.00(08) \\ 
B6  & 0.81(14) & 0.81(08) & 0.14(01) & 0.90(15) & 0.93(15) \\ 

oB6 &  ---     &   ---    & 0.15(02) &  ---     &  ---      \\

\midrule                             
                                      
E5g & 1.02(09) & 1.09(09) & 0.19(02) & 1.08(10) & 1.11(11) \\ 
F6  & 1.00(16) & 0.93(09) & 0.16(02) & 1.04(17) & 1.05(18) \\ 
F7  & 0.72(09) & 0.71(08) & 0.12(01) & 0.78(10) & 0.80(11) \\ 
G8  & 0.58(07) & 0.75(10) & 0.12(02) & 0.60(09) & 0.60(09) \\ 
                                      
\midrule                              
                                      
N6  & 0.67(14) & 0.72(15) & 0.12(02) & 0.68(14) & 0.69(15) \\ 
O7  & 0.40(08) & 0.47(13) & 0.08(02) & 0.41(08) & 0.40(08) \\

\bottomrule
\end{tabular}
\caption{Results of the topological susceptibility
from \eq{eq:chiglob}, with $Q$ measured also at $t=t_1$, 
and \eq{eq:chicorr}. The scales $t_0$ and $t_1$ do always take the value
at the finite quark mass of the corresponding ensemble. 
All the values have been multiplied by a factor $10^3$.
For most of the ensembles, $\chicorr$ turns out to
be more precise. 
\label{tab:topsusc}}
\end{table}

In these figures, it is interesting to note that while the large $x_0$ region shows almost perfect scaling,
a significant cut-off effect is visible at $x_0=0$. Indeed, if  this point is excluded from 
the summation in \eq{eq:chicorr}, the sum does not show any scaling violation beyond 
its statistical accuracy. 

In \tab{tab:topsusc} we also report the result of our ensemble with open BC, namely oB6.
We observe that for this lattice spacing and pion mass the influence of the boundaries in $\langle \bar q(x_0)^2 \rangle $
is visible with the present accuray up to 1.5~fm. Therefore, we restrict the computation
of the susceptibility given in \eq{eq:chicorr} to time slices in the range $[20,171]$;
the summation window remains fixed to $r=5 \sqrt{t_1}$. No differences to B6 are observed in the final result and
the errors are about the same. The reason is that the statistics is roughly the 
same. Furthermore, the reweighting factors of the twisted mass 
reweighting employed in the oB6 ensemble decouple almost completely
from pure gluonic observables such as the charge.

\subsubsection{Numerical estimate of the tail}
\label{s:tail}
We then fitted the correlation function to a single exponential as shown in \fig{fig:chit0},
and estimated the systematic error of $\chi^\mathrm{corr}(r)$ 
from the integral over the fit function. As argued below
this ansatz is justified in this situation despite the four-dimensional smoothing of the flow.

The fit was done to the data with $t=t_0/2$ shown in the figure, 
namely together to the data at the three values of $a$, neglecting scaling violations and in a range 
$2.5\leq x_0\,t_1^{-1/2} \leq 6$. 
The thus determined systematic error turns out to be below 1\%, 
much smaller than
our statistical one. Even if our model underestimated the 
integral by a factor $4$, which seems unlikely given the reasonable
agreement with the data, the tail would still be negligible with our
present accuracy. 

\subsubsection{Effects of the smoothing footprint in the 
large time behavior\label{s:foot}}

The neglected tail gives a systematic error of this measurement. In 
order to justify the above estimate of its contribution, we need to 
be sure that the single exponential ansatz is expected to give a satisfactory
description of the data in the range where it is applied.

The exponential decay with the mass of the flavor singlet
pseudoscalar is modified due to two effects: the contribution of excited states
and the fact that we apply a four dimensional smoothing to our observables. Since
the flow generates a footprint of size $\sqrt{8t}$ with a Gaussian profile,
one expects that for sufficiently large $x_0$ this modification  is negligible.
With a lowest mass $m_0$ in the considered channel, dimensional analysis leads one to expect that it can be neglected once 
$ x_0  / (8t\,m_0) \gg1$.

To gain more insight into the influence of the smoothing 
on the shape of the two-point function, we consider the model of a free smoothed 
scalar field, 
\bes
 \Phi_t(x) \propto \int \rmd^4z\, \rme^{-(z-x)^2/(4t)}\Phi(z)\,,
\ees 
noting that in lowest order of perturbation theory, 
the gluon field is smoothed in exactly the same way. Its time-slice
correlation function at zero spatial momentum is given by
\bes
C_t(x_0) \equiv\int  \rmd\,\mathbf{x}\, \langle \Phi_t(x) \Phi_t(0) \rangle 
 \propto \int \rmd z_0 \, \rme^{-z_0^2/(8t) - m|z_0+x_0|}  \,,
\ees
where $m$ is the the mass of the scalar field. 
The large time asymptotics is given by 
\begin{equation}
C_t(x_0)  
   \propto \rme^{-\mu \xi}\, \left( 
   1 -  \rme^{-(\xi - \mu/2)^2}\,
         \left[\frac{\mu}{2\sqrt{\pi} \xi^2}  
           + \rmO(\xi^{-4}) \right]\right )\quad
\end{equation}
where we used dimensionless variables 
$\mu=\sqrt{8t}\,m\,,\; \xi=x_0/\sqrt{8t}$.
The corrections to the asymptotic exponential decay due to the 
footprint of the smoothing are of order 
$\xi^{-2}\,\rme^{-(\xi - \mu/2)^2}$.
They are small once $\xi - \mu/2 > c$, or equivalently $x_0>c \sqrt{8t}+4tm$,
and $c=1.5$ or $c=3$, with the effect of the smoothing in the per cent and far below the
per mille range, respectively.

While our explicit computation is in a model, we believe that 
it provides a good general guideline. In particular 
the structure with the two different terms is expected to be 
a general feature.

For our case, the relevant mass is $m_0 \approx 1\,$GeV. Choosing 
$t=t_0/2$, we have  $\mu  \approx 1.5$ and $\xi > 2$ translates
into a footprint effect of less than 3\%. 
Since the discussion in \sect{s:tail} is about a systematic effect on the neglected tail, i.e.,
on a small systematic error itself, such an accuracy is more than sufficient
in particular in view of our statistical uncertainties.

\subsection{Distribution of the topological charge}
\begin{figure}
\centering
\includegraphics[width=\textwidth]{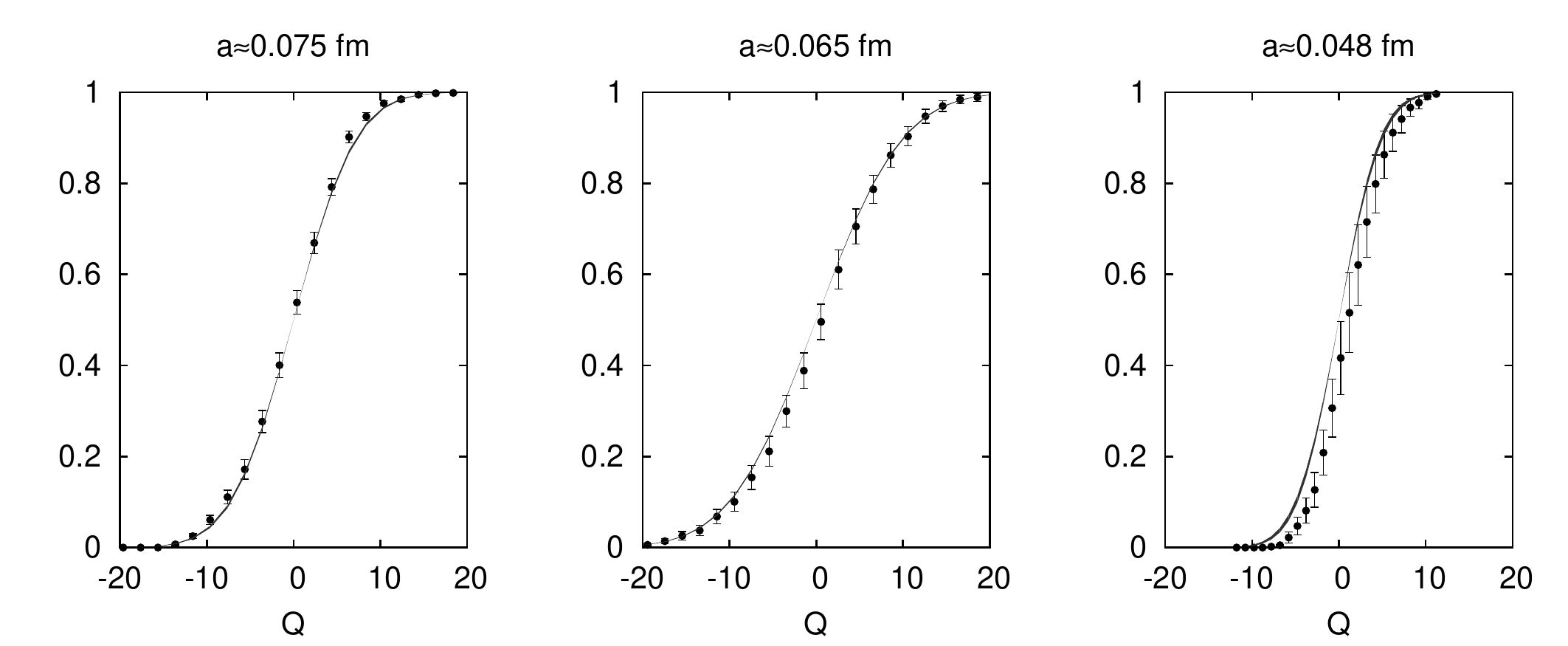}
\caption{Plot of the cumulative distribution function of the topological charge for 
three ensembles. The solid lines represent a Gaussian distribution
with the width coming from the measured topological susceptibility.\label{fig:KS-test}}
\end{figure}

In the large volume limit, the topological charge is expected to follow a 
Gaussian distribution, with the width given by the topological susceptibility.
In \fig{fig:KS-test} we show the cumulative distribution function $C(Q)$, which gives the 
probability of a configuration having charge smaller than $Q$, along with 
the measurement on three different ensembles. The theoretical lines use
as width of the Gaussian distribution the measured value of $\chi^\mathrm{corr}$ of 
\tab{tab:topsusc}. The agreement with a Gaussian distribution is very 
reasonable, with only small deviations visible in the tails at $a=0.048$~fm,
which we attribute to our limited sampling of the tails.

\subsection{Continuum limit and dependence on the light quark mass}
\label{s:cont}
\begin{figure}
\centering
\includegraphics[width=0.7\textwidth]{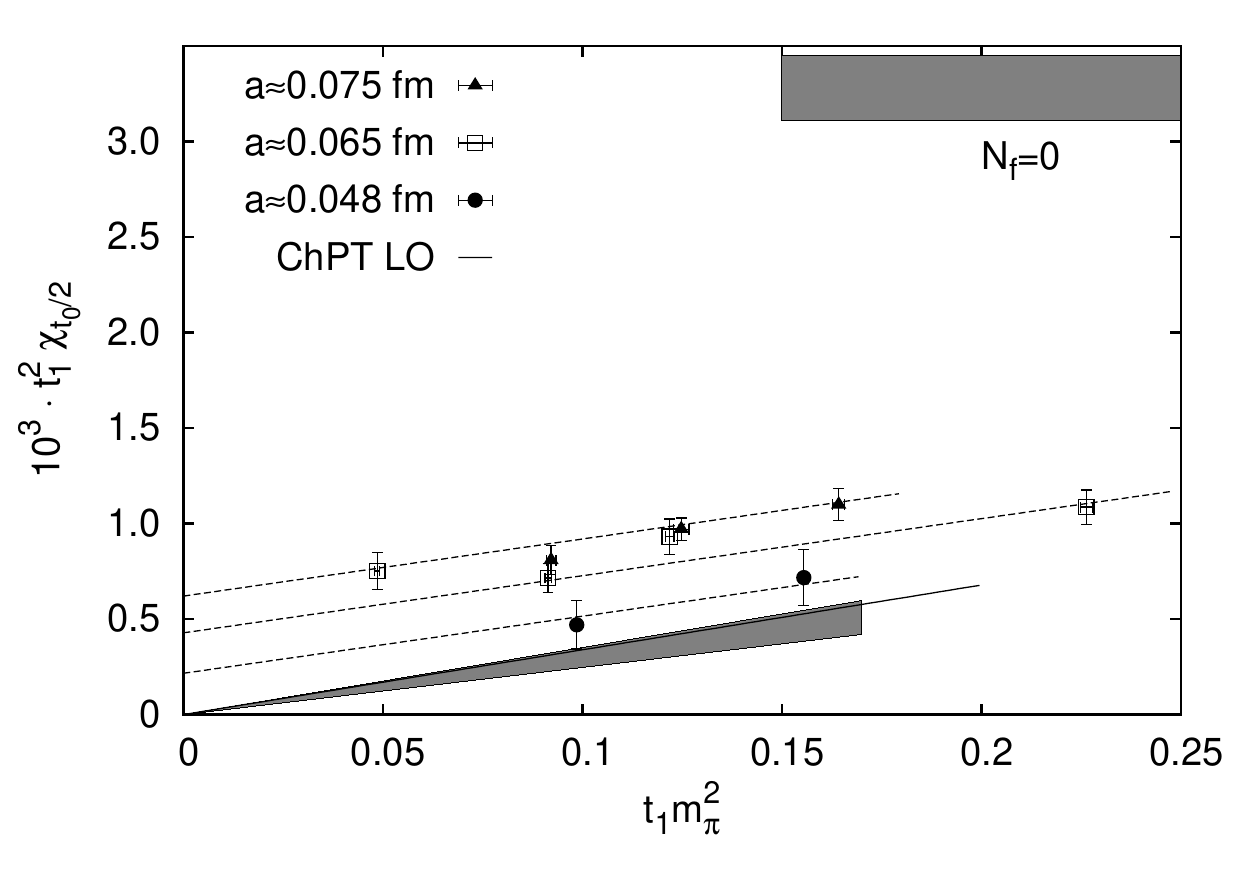}
\caption{The values of the topological susceptibility $\chicorr$ from \tab{tab:topsusc}
along with the leading order expectation from ChPT and the pure gauge value. 
\label{fig:top-susc}}
\end{figure}

In leading order in Chiral Perturbation Theory, the topological susceptibility
is linear in the quark mass\cite{Leutwyler:1992yt}, or using the Gell-Mann--Oakes--Renner
relation
\begin{equation}
\chi = \frac{m}{2} \Sigma \left(1 + O(m)\right) = \frac{1}{8} f_\pi^2 m_\pi^2 (1 + O(m_\pi^2)) \,.
\end{equation}
We therefore display the 
results for $\chicorr$ from \tab{tab:topsusc}  in \fig{fig:top-susc} in 
terms of the dimensionless variable $t_1 m_\pi^2$, where we always take 
$t_1=t_1(m_\pi)$, the 
value of the respective ensemble. With the lattice spacing determined from
the kaon decay constant\cite{Fritzsch:2012wq} and the pion decay constant  set to 130.4~MeV, we obtain the line represented in the figure. Also shown is the 
pure gauge result of \Ref{DelDebbio:2004ns}. The comparison confirms the strong 
effect of the presence of the
fermions found in previous studies.\footnote{Despite the lower statistical accuracy, 
we prefer this result to others obtained with
smearing and cooling because the definition via the index of a chirally symmetric Dirac operator has
a well-defined continuum limit.}

In the figure,  a strong cut-off effect is visible, with the topological susceptibility
significantly enhanced for the coarser lattices. Because of the breaking of chiral symmetry, at finite
lattice spacing there is no reason for the susceptibility to vanish at vanishing pion mass.
Assuming leading effects of $\mathrm{O}(a^2)$, 
Wilson Chiral Perturbation Theory suggests the following ansatz\footnote{We 
are indebted to Oliver B\"ar for communicating unpublished results regarding this issue.}
\begin{equation}
t_1^2 \chitop= c\, t_1\, m_\pi^2+ a^2 \frac{b}{t_1} \,.
\end{equation}
The result is also shown in \fig{fig:top-susc}. As can be seen, the description of the data is satisfactory
given the large error bands and the continuum limit is compatible with the LO ChPT expectation.
We find for the slope $c=2.8(5) \cdot 10^{-3}$, for the intercept
$b=5.1(7) \cdot 10^{-3}$ and a $\chi^2/$d.o.f of the global fit of 1.17.

\section{Conclusions}
Measurements of observables with long autocorrelation times are difficult and the
topological susceptibility is no exception. However, we found that for the action 
used in the CLS simulation, the topological charge is not yet the slowest observable
in the range of lattice spacings under investigation, though the rapid growth in $\tauint$
indicates that on slightly finer lattices the charge would effectively freeze in typical
simulations. We show that light quarks mitigate the problem of the slow topological charge
to a certain extent, whereas the other slow modes, as seen, e.g., in  $E(t_0)$, do not
profit.

Larger quark masses and also cut-off effects (for the action employed in this paper) both
increase the susceptibility, however, even on our coarsest lattices and largest pion mass of 
$630$\,MeV, we observe a suppression of the susceptibility with respect to the pure gauge
case by more than 50\%. The three lattice spacings with a range of pion masses at our disposal
allow us to take the continuum limit with theoretical support from 
Wilson Chiral Perturbation Theory. As we have shown, for fitting the pion mass dependence 
of $\chitop$ at finite lattice spacing, it is crucial to include a term
that does not vanish at zero pion mass and finite cutoff. Only with chirally symmetric
sea quarks such a term can be excluded.

We stress that discretization effects seem large, 
because the quantity of interest is small for small quark masses. 
In such a situation,
large relative discretization errors are generic, however, the scale to judge them is the
deviation from the pure gauge theory result and on this scale we reach a percent level accuracy
in the continuum for the whole quark mass region.

Finally let us point out that the discussion in \sect{s:foot} 
on the effects of
the footprint is particularly relevant
when one wants to extract masses from correlation functions at
positive flow time. For example the glueball mass determination in 
\cite{Chowdhury:2014kfa} may be affected.

\subsection*{Acknowledgments}
We thank Francesco Virotta for collaboration in an early stage of this project.
It is also a pleasure to thank Oliver B\"ar for essential discussion concerning
the Wilson Chiral Perturbation Theory prediction and Stefan Sint and Ulli Wolff for discussion.
This project used  $N_f=2$ gauge
field configurations generated within the CLS effort; we are grateful to our colleagues for sharing them with us. 
We thankfully acknowledge the computer resources
granted by the John von Neumann Institute for Computing (NIC)
and provided on the supercomputer JUROPA at J\"ulich
Supercomputing Centre (JSC) and by the Gauss Centre for
Supercomputing (GCS) through the NIC on the GCS share
of the supercomputer JUQUEEN at JSC,
with funding by the German Federal Ministry of Education and Research
(BMBF) and the German State Ministries for Research
of Baden-W\"urttemberg (MWK), Bayern (StMWFK) and
Nordrhein-Westfalen (MIWF), as well as
within the Distributed European Computing Initiative by the 
PRACE-2IP, with funding from the European Community's Seventh 
Framework Programme (FP7/2007-2013) under grant agreement RI-283493,
by the HLRN in Berlin, and by NIC at DESY, Zeuthen.

\bibliography{topo}
\bibliographystyle{JHEP}

\end{document}